\documentclass[prl, twocolumn, superscriptaddress]{revtex4-1}

\usepackage{graphicx}

\begin{document}

\title{Three-dimensional Imaging of Cavity Vacuum with Single Atoms Localized by a Nanohole Array}

\author{Moonjoo Lee}
\altaffiliation{Present address: Institute for Quantum Electronics, ETH Z\"urich, CH-8093 Z\"urich, Switzerland}
\author{Junki Kim}
\author{Wontaek Seo}
\author{Hyun-Gue Hong}
\author{Younghoon Song}
\affiliation{Department of Physics and Astronomy, Seoul National University, Seoul 151-747, Korea} 
\author{Ramachandra R.\ Dasari}
\affiliation{G.\ R.\ Harrison Spectroscopy Laboratory, Massachusetts Institute of Technology, Cambridge, MA 02139, U.S.A.}
\author{Kyungwon An}
\email{kwan@phya.snu.ac.kr}
\affiliation{Department of Physics and Astronomy, Seoul National University, Seoul 151-747, Korea} 
\date{\today}

\begin{abstract}
Zero-point electromagnetic fields were first introduced in order to explain the origin of atomic spontaneous emission. 
Vacuum fluctuations associated with the zero-point energy in cavities are now utilized in quantum devices such as single-photon sources, quantum memories, switches and network nodes.
Here we present 3D imaging of vacuum fluctuations in a high-$Q$ cavity based on the measurement of position-dependent emission of single atoms. 
Atomic position localization is achieved by using a nanoscale atomic beam aperture scannable in front of the cavity mode. 
The 3D structure of the cavity vacuum is reconstructed from the cavity output. 
The rms amplitude of the vacuum field at the antinode is also measured to be $0.92\pm0.07$ V/cm.
The present work utilizing a single atom as a probe for sub-wavelength imaging demonstrates the utility of nanometer-scale technology in cavity quantum electrodynamics.
\end{abstract}
\maketitle

\section{Introduction}
Single nanoscopic particles such as atoms \cite{Hood-Science00}, molecules \cite{Vollmer-NatMethods08, Ransik-NatMethods06, Cang-Nature11} and quantum dots \cite{Ropp-NatComm13} are an ideal probe
for sub-wavelength imaging applications. A single atom, in particular, is an ultimate detector for
electromagnetic (EM) fields owing to its sub-nanometer size and strong interaction with light \cite{Streed-NatComm12}. 
For sub-wavelength imaging with single atoms, it is then necessary to localize the atoms in the nanometer scale.
Single ions confined in space with specially arranged electrodes have been used as a probe 
for mapping the transverse and axial profile of  an intracavity EM field \cite{Guthohrlein-Nature01, Steiner-PRL13}.

Implementing such localization with single neutral atoms is rather difficult in contrast to single ions. 
Nonetheless, a single neutral atom was tightly localized in an optical lattice in the Lamb-Dicke regime \cite{Bloch-RMP10, Boyd-Science06} and its spectrum was observed to determine its inter-potential-well tunneling rate \cite{Kim-NL10}. 
An intracavity dipole trap was used to confine a single neutral atom near an antinode for maximum atom-cavity interaction while 
heating effects induced uncertain atomic delocalization \cite{Maunz-PRL05}. 
Recently, three-dimensional sideband cooling \cite{Reiserer-PRL13} was demonstrated to localize single atoms tightly in a cavity. However, it was not yet possible to scan their positions continuously in the nanometer scale 
for advanced photonics applications using the atom-cavity coupling constant as a variable \cite{Dubin-NPhys10, Choi-PRL10}.

Vacuum fluctuations, the subject of imaging in the present study, arise from the creation and annihilation of virtual particles. 
Macroscopically, the vacuum energy density associated with the vacuum fluctuations is known to be proportional to the cosmological constant \cite{Weinberg-RMP89}. 
Microscopically, vacuum fluctuations lead to the Lamb shift \cite{Fragner-Science08} and the spontaneous emission of atoms. 
Moreover, the Casimir effect \cite{Casimir-PKNAW48} was observed between two metallic plates or dielectric mirrors.
The cavity vacuum, coupled with single atoms \cite{Mckeever-Science04, Wilk-Science07, Ritter-Nature12}, superconducting qubits \cite{Fragner-Science08}, cold atomic ensembles \cite{Tanji-Suzuki-Science11, Bohnet-Nature12} and Bose-Einstein condensates \cite{Brennecke-Science08}, makes it possible to perform a wide range of fundamental studies based on its spatial structure \cite{Hood-Science00, Haroche-EPL91, Daul-EPJD05} and amplified strength \cite{An-PRL94, Choi-PRL06, Hong-PRL12}. 

Although vacuum field has no classical counterpart, it is common practice to assume its spatial structure based on the solution of Maxwell's equations \cite{Casimir-PKNAW48, Raimond-RMP01, Hood-Science00}. 
The rms amplitude of a vacuum field is then obtained from zero-point energy in the quantum electrodynamics (QED) theory.
Under field quantization in the absence of dissipation \cite{Milonni-Vacuum94}, the root mean square (rms) vacuum field corresponding to a TEM$_{00}$ mode of a Fabry-P\'{e}rot-type cavity is given by 
\begin{equation}
E_{\rm vac}(x,y,z)=\sqrt{\frac{\hbar\omega}{2\epsilon_{0}V}} e^{[-(x^2+y^2)/w_{0}^2]}\cos{\left(\frac{2\pi z}{\lambda}\right)},
\label{eq1}
\end{equation}
when the mirror spacing is much less than the Rayleigh range. Here
$\epsilon_0$ is the permittivity of vacuum, $V$ is the mode volume, $w_0$ is the cavity mode waist, and $\lambda(\omega)$ is the resonant wavelength (angular frequency). 
An excited two-level atom resonant with this vacuum field would then undergo a vacuum-field-driven transition, the rate of which for a short interaction time is proportional to the vacuum intensity $E_{\rm vac}^2(x,y,z)$. 

Previously, the spatial structure of the vacuum field in a cavity was partially revealed with single trapped ions under more dominant interactions with free-space vacuum \cite{Kreuter-PRL04}.
The rms amplitude of the vacuum field was also obtained from measurements of vacuum Rabi frequency 2$g(\mathbf{r})\equiv 2\mu E_{\rm{vac}} (\mathbf{r}) / \hbar$ with $\mu$ the atomic induced dipole moment \cite{Raimond-RMP01}. 
However, all of these measurements were either limited to a fixed atomic position with uncertain delocalization effects \cite{Maunz-PRL05} and unwanted displacement of the atomic equilibrium position \cite{Steiner-PRL13} or averaged over several wavelengths \cite{Thompson-PRL92}.

\begin{figure*}
\includegraphics[width=4.5in]{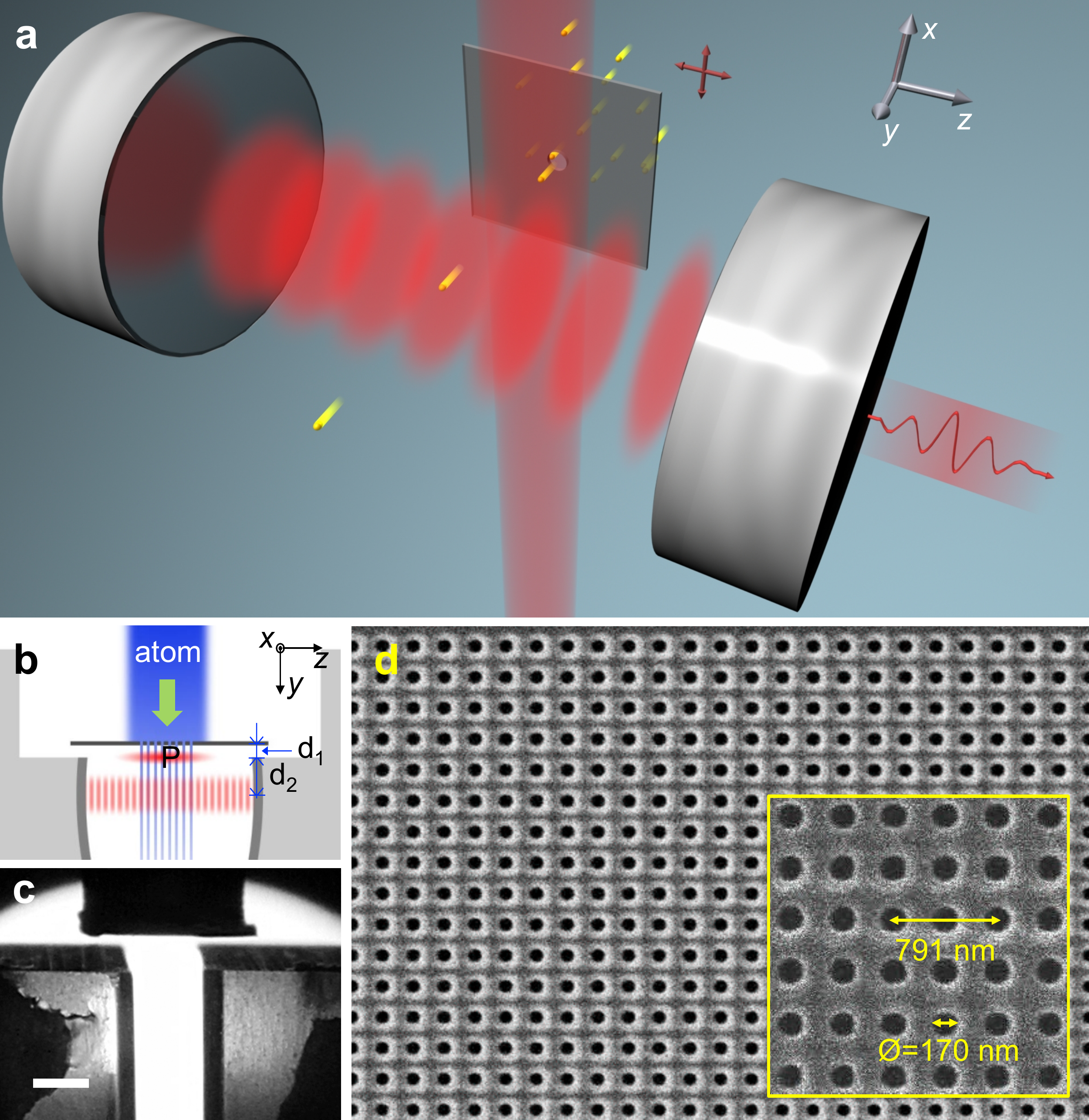}
\caption{Imaging vacuum fluctuations in a cavity with single atoms and a nanohole array.
(a) An ideal scheme.
A fast-moving single two-level atom, after passing through a nanohole, is excited by a pump laser travelling in the $-x$ direction with its polarization in the $y$ direction. 
Photons emitted out the cavity, proportional to the intensity of the vacuum fluctuations, are measured for various $(x,z)$ positions of the nanohole aperture.
Information along the $y$ axis is obtained from transit-time broadened atom-cavity scan curves. 
(b) Experimental setup. 
An aperture having an array of nanoholes with a period of a half of the resonant wavelength is employed. 
To overcome delocalization of atoms due to a finite atomic beam divergence, 
the cavity substrate is machined (see Methods, \cite{Kim-OL12}) to allow the nanometric aperture to approach the cavity mode as closely as possible. 
Here, $d_{1}\sim 120$ $\mu$m and $d_{2}\sim 300$ $\mu$m. $P$ denotes the pump laser and the $y$ axis is the quantization axis.  
(c) Photo of the cavity and the approached nanohole array. The scale bar indicates a length of 1 mm.
(d) FIB image of the nanoholes. Pitches are $0.5\lambda$, and the diameter is about 170 nm (0.2$\lambda$).}
\label{fig1}
\end{figure*}

Here we introduce a nanohole array \cite{Kolbel-NL2002} to control the atomic position and use single atoms as a nanoscopic probe for the vacuum field in the cavity. 
Pre-excited single atoms in the cavity respond to the cavity vacuum field and emit single photons to the cavity mode. 
We measure the photons emitted out the cavity.
The number of photons is then proportional to the vacuum-field intensity. 
We map out the vacuum intensity distribution by scanning the nanohole array along two transverse directions. 
The vacuum  intensity profile along the direction of atomic beam is obtained from the transit-time broadening 
in the atom-cavity detuning dependence of the observed signal. 
In this way the 3D profile of the vacuum-field intensity is imaged with a spatial resolution of about 170 nm, 
which is mostly limited by the nanohole diameter.

\section{Result}

\noindent\textbf{Design of experiment.}
An ideal experimental scheme to image the vacuum-field intensity $E_{\rm vac}^2(x,y,z)$ of Eq.~(\ref{eq1}) is sketched in Fig.~\ref{fig1}.
A key element is a nanohole atomic beam aperture, whose hole diameter is much smaller than $\lambda$. 
Atomic position localization and short interaction time are achieved by injecting excited two-level atoms one by one at a large velocity through the nanohole aperture. The aperture position is scanned along the axial ($z$) and transverse ($x$) directions just in front of the cavity mode.
The interaction time $\tau$ between the atom and the cavity mode is so brief that the vacuum Rabi angle $2g(\mathbf{r})\tau$ is much smaller than unity.
Another requirement is that the injected atom should interact mostly with the cavity vacuum. 
Its free-space emission should be negligible during the transit through the cavity and
the photon emitted by a preceding atom should decay the cavity well before the entry of the next atom.
The latter is equivalent to the condition that the intracavity mean photon number $\langle n \rangle$ in the steady state should be much less than unity.
The atomic emission rate, and thus $\langle n \rangle$, will then be proportional to the intensity $E_{\rm vac}^2$ of the zero-point electromagnetic field fluctuating in the single cavity mode. 
Variation of $E_{\rm vac}^2$ along the atomic path (in the $y$ direction) is encoded in the transit-time broadened atom-cavity scan curve of $\langle n \rangle$ 
\cite{Demtroder-LS02}.
Combining this information, we can reconstruct the 3D structure of the vacuum field.

The above experimental scheme performed in the {\em linear} regime ({\em i.e.}, negligible Rabi angle and $\langle n \rangle \ll1$) does not provide the actual rms amplitude $E_{\rm{vac}}(\mathbf{0})$ of the vacuum field. 
However, the rms amplitude can be obtained from measurements in the {\em nonlinear} regime ({\em i.e.}, appreciable Rabi angle and $\langle n \rangle$ comparable to or larger than unity), corresponding to the emergence of one-atom lasing \cite{Meschede-PRL85, An-PRL94}. 
This will be explained below and also in Ref.~\cite{supp}.

In our actual experiment depicted in Fig.~\ref{fig1}(b)-(d), we use a Fabry-P\'{e}rot type cavity made of two mirrors with a 10-cm radius of curvature and a 1.09 mm mirror spacing. The resulting Rayleigh range is 7.4 mm and thus the mode waist $w_0$ is expected to remain almost constant at 43 $\mu$m (the largest variation less than 0.1 $\mu$m) along the entire axial ($z$) range of the mode.
We employ a supersonic beam of $^{138}$Ba atoms and utilize its $^{1}$S$_{0}\leftrightarrow^{3}$P$_{1}$ transition with $\lambda$=791.1 nm and $\gamma=2\pi\times50$ kHz, the free-space spontaneous emission rate. 
The maximum atom-cavity coupling constant $g_0=\mu E_{\rm vac}(\mathbf{0})/\hbar$ is expected to be $2\pi \times 330$ kHz, which surpasses $\gamma$ and the cavity decay rate $\kappa=2\pi\times140$ kHz, thus putting our experiment in the strong coupling regime.  The cooperativity parameter $g_0^2/\gamma\kappa$ is $16$.

Although only one nanohole is assumed in the above ideal scheme, in actual experiments it is impossible to have an intracavity mean atom number $\langle N \rangle$ sufficient to yield a reasonable signal-to-noise ratio for $\langle n \rangle$ even at the highest atomic beam flux when only one nanohole is used. 
Instead, we employ an array of nanoholes with a period of $\lambda/2$  [Fig.~\ref{fig1}(d)] in order to increase $\langle N \rangle$ to a practical level ranging from 0.1 to 1.5. 
This approach is acceptable since the vacuum field of interest is expected to be periodic along the cavity axis, based on the mode solution of Maxwell's equations. 
It works even for an arbitrary cavity vacuum via Fourier transform as discussed below and in Ref.~\cite{supp}.
The nanoholes with a diameter of 170 nm were milled with the focused ion beam (FIB) technique on a 75-nm-thick silicon nitride membrane [Fig.~\ref{fig1}(d)]. 
The total number of holes is 72 $\times$ 16 in the $z \times x$ directions, spanning a range of 28.5 $\mu$m $\times$ 6.3 $\mu$m on the membrane. 
Since the $x$ direction span range is much smaller than the mode waist $w_0$, we have enough imaging resolution in that direction.\\

\begin{figure*}
\includegraphics[width=5in]{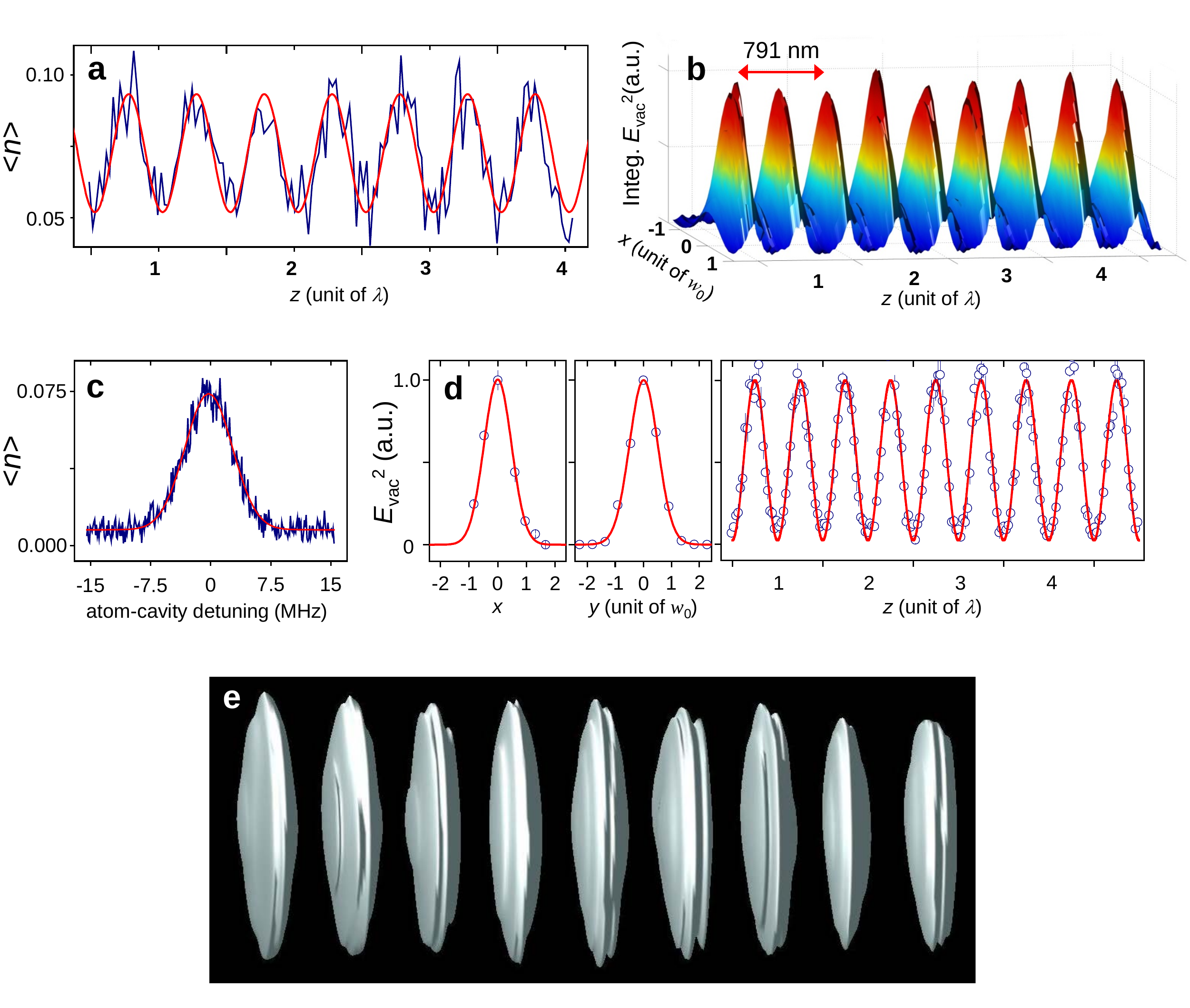}
\caption{3D image of the cavity vacuum-field intensity.
(a) Intracavity mean photon number (in blue) as a function of the $z$ coordinate of the nanohole array when its $x$ coordinate is zero, at the center of the cavity mode. Also shown is the theoretical curve (in red), the result of the quantum master equation when $\langle N \rangle = 0.34$. 
(b) Result of deconvolution of the raw data in the $xz$ plane with the atomic spatial distribution as a point spread function. 
The result represents $\int E_{\rm vac}^2(x,y,z)dy$, an integral along the atomic path. 
(c) Transit-time broadened atom-cavity detuning curve with the nanoholes being fixed. The red line is a Gaussian fit. The Fourier transform of its amplitude reflects the $y$-direction profile of the cavity vacuum.
(d) Slice cuts of the reconstructed vacuum-field intensity $E_{\rm vac}^2(x,y,z)$ along the $x$, $y$ and $z$ axes. 
Red curves are Gaussian fits along the $x$ and $y$ axes and a sine-square fit along the $z$ axis. 
(e) An iso-surface of $[E_{\rm vac}(x,y,z)/E_{\rm vac}(\mathbf{0})]^2=0.2$, visualizing the 3D structure of the vacuum fluctuations in the cavity mode.
}
\label{fig2}
\end{figure*}

\noindent\textbf{3D imaging of cavity vacuum.}
The trace in Fig.~\ref{fig2}(a) shows the observed $\langle n \rangle$ as a function of the $z$ coordinate of the nanohole array while its $x$ coordinate is fixed near the mode center ($x=0$). We keep the cavity on resonance with a feed-forward technique \cite{supp}. The oscillatory feature comes about since the atomic emission rate is enhanced/suppressed when the nanohole columns are aligned with the antinodes/nodes of the cavity vacuum. 
The data in Fig.~\ref{fig2} were obtained with an optimal value of $\langle N\rangle(=0.34)$ which simultaneously maximize the signal-to-noise and the probability of having zero photons in the cavity when each atom arrives at the mode.
The probability is more than 91\%, and therefore we can attribute the oscillatory feature mostly to the spatial variation of the cavity vacuum.
The vacuum Rabi angle $2g_0\tau$ acquired during the entire transit time along an antinode is only $0.12\pi$, justifying that the data were taken in the linear regime. 
Any deviation from the linearity will show up in the fitting error \cite{supp}.
The observed $\langle n \rangle$ is in good agreement with the theory curve based on the quantum master equation \cite{Filipowicz-PRA86, An-PRL94} with a period of $\lambda/2$. 

The limited visibility in Fig.~\ref{fig2}(a) is due to the atomic position spread after the nanoholes. 
In order to reveal the latent structure of the vacuum field, we deconvolved the raw data while regarding the atomic spatial distribution as a point-spread function. 
The Richardson-Lucy algorithm \cite{Richardson-JOSA72} was used for deconvolution. 
Combining similar traces measured for 7 different $x$ coordinates at a 15-$\mu$m interval, we then  obtain the curve in Fig.~\ref{fig2}(b), which is proportional to an integrated $E_{\rm vac}^{2}$ over the $y$ coordinate when an atom is injected onto a point in the $xz$ plane. 
Fitting the trace for each $x$ coordinate with a sine-squared function in $z$ yields a resonant wavelength of $\lambda$=791.0$\pm$0.6 nm, consistent with the known transition wavelength of barium.

The temporal profile of the vacuum field along the $y$ direction during the atomic transit is obtained by Fourier-transforming the amplitude associated with the transit-time broadened atom-cavity scan curve in Fig.~\ref{fig2}(c) \cite{Demtroder-LS02}, which was taken with the aperture position fixed. Using the atomic velocity of 830 m/s independently measured with Doppler spectroscopy, we obtain the spatial profile of the vacuum field along the $y$ direction. 
Combining the information from all three axes, we can reconstruct the 3D structure of the vacuum-field intensity as shown in Fig.~\ref{fig2}(d). 
The resulting  profile in the $x$($y$) direction is a Gaussian with a mode waist of 41$\pm$2 $\mu$m (43$\pm$1 $\mu$m), agreeing well with the expectations from the cavity geometry within the experimental uncertainty.
An iso-surface of $[E_{\rm vac}(x,y,z)/E_{\rm vac}(\mathbf{0})]^2=0.2$ is also shown in Fig.~\ref{fig2}(e).\\

\begin{figure}
\includegraphics[width=3.4in]{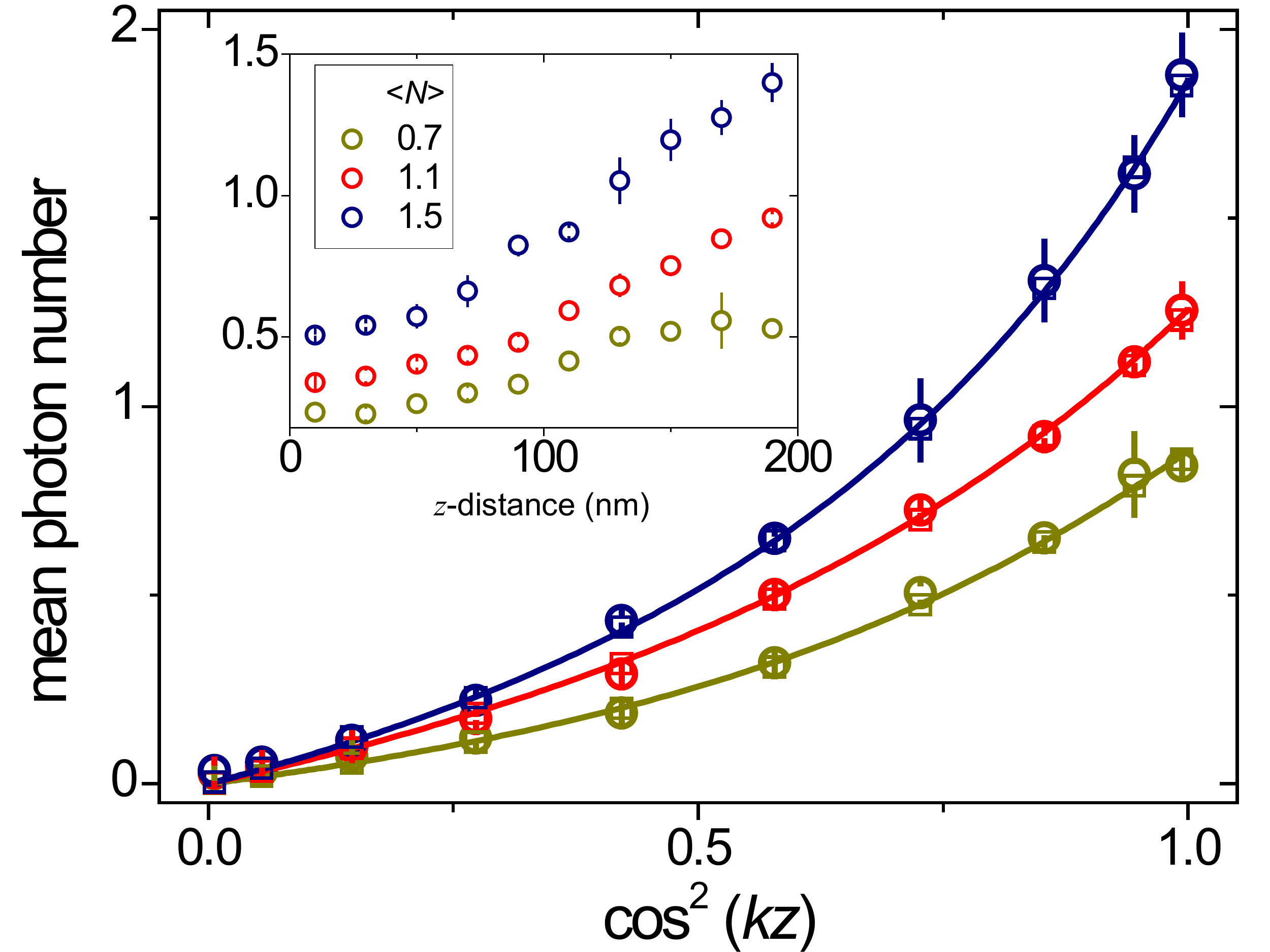}
\caption{Numerical fit to yield the amplitude of the vacuum field.
Deconvolved mean photon number $\langle n \rangle$ (circle) as a function of the relative vacuum intensity $[E_{\rm vac}(0,0,z)/E_{\rm vac}(0,0,0)]^2$ for three different values of $\langle N \rangle$. 
Note that the nonlinearity becomes more pronounced as $\langle N \rangle$ gets larger than unity.
The inset shows the raw data of $\langle n \rangle$  as the $z$ coordinate of the nanohole array changes from a node to an antinode.
Data fitting was performed for $\langle N \rangle$=1.1 and 1.5 with the solution of the quantum master equation.
The rms amplitude $E_{\rm{vac}}(\mathbf{0})$ was then uniquely determined to be $0.92\pm0.07$ V/cm by the fitting. 
The quantum trajectory simulation (square), which correctly treats multi-atom events in the cavity, confirms that the fitting based on the master equation is valid even when $\langle N \rangle$ is larger than unity. 
The background offset was subtracted from the raw data before deconvolution. The error bar is the sum of the shot
noise and the numerical error in the deconvolution process \cite{supp}.
}
\label{fig3}
\end{figure}

\noindent\textbf{The rms amplitude of the cavity vacuum field.}
We have also obtained the actual rms amplitude $E_{\rm vac}({\bf 0})$ of the vacuum field from an independent experiment in the nonlinear regime.
The cavity output was recorded as the nanoholes was scanned along $z$ axis by $\lambda/4$, from a node to an antinode, for 3 different $\langle N \rangle$ (Fig.~\ref{fig3} inset). 
The amplitude $E_{\rm vac}({\bf 0})$ was then obtained by fitting the nonlinear data in Fig.~\ref{fig3} with the solution of the master equation with $E_{\rm vac}(\mathbf{0}), \langle n \rangle$ and $\langle N \rangle$ as fitting parameters \cite{supp}. 
Since the data in Fig.~\ref{fig3} contain nonlinear interactions between the atom and the cavity mode as a result of appreciable $\langle n \rangle \sim 1$, the three parameters are uniquely determined by the fit. 
All of the $\langle n \rangle$ and $\langle N \rangle$ values appearing in Fig.~\ref{fig2} and Fig.~\ref{fig3} have been calibrated with this method.
The fit also gives the rms amplitude $E_{\rm vac}(\mathbf{0})$=0.92$\pm$0.07 V/cm.

\section{Discussion}

The cavity mode volume $V$ in Eq.~(\ref{eq1}) can be obtained by spatially integrating the absolute square of 
the mode function $f(\mathbf{r})=E_{\rm vac}(\mathbf{r})/E_{\rm vac}(\mathbf{0})$.
The mode function is given by the $w_0$ and $\lambda$ values obtained from our 3D imaging experiment and  
the mirror spacing measured from the microscope image of the cavity \cite{Kim-OL12}. 
The integration yields a mode volume of 1.5$\pm$0.1 nanoliter. 
This value is in good agreement with a dissipation-free mode volume $V_0$ of 1.52$\pm$0.04 nanoliter calculated from the cavity geometry, 
{\em i.e.}, the radius of curvature of mirrors and the mirror spacing. 
The observed mode volume then yields a dissipation-free rms amplitude $E_{\rm vac}^{\rm (df)}=\sqrt{\hbar \omega/2\epsilon_0 V}$ of the vacuum field as 0.97$\pm$0.03 V/cm, based on the assumption of  zero-point energy conservation \cite{Senitzky-PR60}.

This value is in good agreement with the rms amplitude $E_{\rm vac}({\bf 0})$ obtained from the above nonlinear regime experiment.
The small discrepancy may draw one's attention to the long-standing problem of the field quantization in open systems, related to the sustainability of the zero-point energy and the uncertainty principle under dissipations \cite{Dekker-PR81}. However, the current experimental accuracy and precision does not mandate such considerations.


Although a specific cavity vacuum field is investigated in the present work, our approach can provide complete imaging of an {\em arbitrary} cavity vacuum via Fourier transform.
The cavity output signal obtained as a function of the nanohole array position like the trace in Fig.~\ref{fig2}(a) is the convolution of the vacuum intensity distribution and the response function of the nanohole array.
We can then use the convolution theorem to obtain the Fourier transform of the vacuum intensity distribution as the ratio of Fourier transform of the convoluted trace to that of the response function \cite{supp}.
This approach works in the imaging in the $x$-$z$ plane orthogonal to the atomic beam direction.
An arbitrary longitudinal ($y$ direction) profile can be extracted from the transit-time broadened atom-cavity scan curves taken at various atomic incidence angles in the $x$-$y$ plane.

In summary, we have performed 3D imaging of the vacuum-field intensity in a cavity by using single two-level atoms as a nanoscopic probe and a nanohole array for atomic position localization. We have also measured the rms amplitude of the vacuum field from the nonlinear regime experiment.
Any other vacuum field mode with a different frequency can in principle also be imaged by choosing a different atom resonant with it except for very short wavelengths at which the mirrors do not function properly \cite{Casimir-PKNAW48}. 
Moreover, the periodic nanohole array enables precise control of the atom-cavity coupling constant in cavity QED studies.
One application might be controlled phase imprinting on atoms before they enter the cavity mode, which would be essential for exciting a bi-phase field or a Schr\"{o}dinger cat state in the cavity \cite{Kim-ICAP12-Nature}.

\section{Methods}
\noindent\textbf{Nanohole array}
The nanohole array was made of a 75-nm-thick silicon nitride membrane (250 $\mu$m $\times$ 250 $\mu$m) with each hole with a diameter of 170 nm milled with the FIB technique (FIB200 by FEI).
Typical milling current and voltage were 1.0 nA and 30.0 kV, respectively. It took about 30 seconds for fabrication of 72$\times$16 nanoholes.
To obtain FIB images, as in Fig.\ 1(d), of nanoholes without damaging them, FIB scanning with 10.0 pA / 30.0 kV was appropriate.

For the experiments done in the linear regime, we used an aperture with 72 $\times$ 16 holes in the $z \times x$ directions, spanning a range of 28.5 $\mu$m $\times$ 6.3 $\mu$m on the membrane. 
For the nonlinear-regime experiment requiring $\langle N \rangle$ to be more than unity, we used a nanohole array with the $x$ range of nanoholes extended twice, i.e., 72 $\times$ 32 holes with a hole diameter of 230 nm.
The nanohole array assembly was mounted on a nanometre-precision positioner and its $z$ coordinate was tracked in real time by a Michelson interferometer with one arm attached on the aperture mount. \\

\noindent\textbf{Cavity modification}
The de Broglie wavelength of the atoms injected into the cavity is $3.4\pm0.3$ pm, which guarantees that atomic matter-wave diffraction by the nanoholes is negligible. 
However, the atomic beam itself has a small but finite angular divergence of 0.24 mrad. Unless the nanohole array is placed within 500 $\mu$m from the cavity mode axis, atomic position localization with a reasonable contrast cannot be achieved.
We solved this problem by grinding one side of the mirror (7.75 mm in diameter and 8 mm in length) with a diamond wheel in the shape of `L' down to $300$ $\mu$m from the mode without damaging the remaining mirror surface [see Fig.\ 1(b)-(c)] \cite{Kim-OL12}.
During the grinding, the mirror surface was covered with a polymer coat for protection.
We found the cavity finesse unchanged from its original value of $1.0\times10^6$.

\newpage
\hbox{}
\hbox{}

\noindent\textbf{Acknowledgement}
We thank H.\ Nha, Y.\ Chough, Y.-I.\ Shin, H.\ Jeong and W.\ Jhe for helpful discussions.
This work was supported by the Korea Research Foundation (Grant Nos.\ 20110015720 and 2013R1A1A2010821).\\

\noindent\textbf{Author Contributions}
M.L. and K.A. designed the experiment. M.L., J.K., W.S., H.H. and Y.S. performed the experiment. M.L. analyzed the data and carried out theoretical investigations. K.A. supervised overall experimental and theoretical works. M.L. and K.A. wrote the manuscript. All authors participated in discussions.

\end{document}